\documentclass[a4paper,11pt]{article}
\usepackage[english]{babel}
\usepackage{mathtools}
\usepackage{pos}
\usepackage{lipsum}
\usepackage{wrapfig}
\usepackage{csquotes}
\usepackage{caption}
\usepackage{subcaption}
\usepackage{siunitx}
\usepackage{titlesec}
\usepackage{hyphenat}
\usepackage{tikz}
\usetikzlibrary{positioning}
\usepackage{cleveref}

\title{PineAPPL Grids of Open Heavy-Flavor Production in the GM-VFNS}

\author*[a]{Jan Wissmann}
\emailAdd{jan.wissmann@uni-muenster.de}
\author[a]{Tomáš Ježo}
\author[b]{Ingo Schienbein}
\author[c]{Hubert Spiesberger}
\author[a]{Michael Klasen}

\affiliation[a]{Institut für Theoretische Physik, Universität Münster,\\
Wilhelm-Klemm-Straße 9, Münster, Germany}

\affiliation[b]{Laboratoire de Physique Subatomique et de Cosmologie, Université Grenoble-Alpes,\\
CNRS/IN2P3, 53 Avenue des Martyrs, 38026 Grenoble, France}

\affiliation[c]{PRISMA+ Cluster of Excellence,
Institute for Nuclear Physics and Institute of Physics,\\
Johannes Gutenberg-University, Staudingerweg 9, 55099 Mainz, Germany}

\abstract{Many next-to-leading order QCD predictions are available through Monte Carlo (MC) simulations. Usually, multiple CPU hours are needed to calculate predictions at a required precision, which is unfeasible for global PDF analyses. This problem is solved by a process known as \emph{gridding}: The values of the hard-scattering cross-section are calculated only once with the MC program, and then interpolated and stored in look-up tables (grids) of the kinematical variables. To obtain the physical predictions, they are convolved with the PDFs (e.g.~during the fitting stage in a PDF global analysis), which takes a tiny fraction of the time needed to calculate the MC results. This is possible with PineAPPL, a library tackling the aforementioned process of grid creation and convolution. In this work, we use PineAPPL to grid the predictions for open heavy-flavor production in the general-mass variable-flavor-number scheme (GM-VFNS). In the GM-VFNS, the differential cross-section interpolates between the fixed-flavor-number scheme (FFNS) and the zero-mass variable-flavor-number scheme (ZM-VFNS). These are each only valid in different kinematical regions, in which the GM-VFNS cross-section reproduces the FFNS and ZM-VFNS as the limiting cases of high energies and small masses, respectively. Better than permille agreement is achieved between the grids and the MC predictions, while at the same time not substantially increasing the time of the MC calculations.}

\FullConference{31st International Workshop on Deep Inelastic Scattering (DIS2024)\\
8–12 April 2024\\
Grenoble, France\\}

\hypersetup{
    pdfauthor={Jan Wissmann},
    pdftitle={PineAPPL Grids of Open Heavy-Flavor Production in the GM-VFNS},
    pdfsubject={DIS2024 Proceedings}
}

\NewDocumentCommand{\m}{m}{m_{\rm #1}}
\NewDocumentCommand{\mQ}{}{\m{Q}}

\NewDocumentCommand{\pT}{}{p_{\rm T}}

\NewDocumentCommand{\alphas}{}{\alpha_{\rm s}}
\RenewDocumentCommand{\d}{}{\mathrm{d}}

\NewDocumentCommand{\comment}{m}{\iffalse#1\fi}

\makeatletter
\newcommand*{\centerfloat}{%
  \parindent \z@
  \leftskip \z@ \@plus 1fil \@minus \textwidth
  \rightskip\leftskip
  \parfillskip \z@skip}
\makeatother

\titlespacing{\section}{0pt}{4pt plus 4pt minus 2pt}{4pt plus 2pt minus 2pt}

\renewcommand{\hookAfterAbstract}{%
\begin{tikzpicture}[remember picture,overlay]
    \node [below left=5mm and 4mm of current page.north east] (preprint) {MS-TP-24-17};
\end{tikzpicture}%
}

\begin{document}
\maketitle

\section{Introduction}

The heavy-quark production process plays an important role in constraining the gluon PDF at low parton momentum fraction \(x\) \cite{duwentaster_impact_2022}.
Several procedures to calculate theoretical predictions for this process exist, such as fixed flavor-number schemes (FFNS) \cite{mangano_heavy-quark_1992}, FONLL \cite{cacciari_p_t_1998}, the Crystal Ball approach \cite{kom_pair_2011} (which was used in \cite{duwentaster_impact_2022} and involves fitting matrix elements to data), and general-mass variable-flavor-number schemes (GM-VFNS) \cite{kniehl_inclusive_2005,kniehl_collinear_2005}.
To use these in PDF global analyses, calculating them must be highly performant, as can be achieved by fast convolution grids \cite{kluge_fastnlo_2007,carli_posteriori_2010,carrazza_pineappl_2020}.
However, no automated GM-VFNS predictions nor grids for these processes are publicly available yet.

\section{A GM-VFNS for Heavy-Flavor Hadroproduction}

The GM-VFNS used in this work deals with the production of a heavy meson (such as a D- or B-meson) in hadron-hadron collisions.
It combines two schemes: The FFNS (fixed-flavor number scheme) and the ZM-VFNS (zero-mass variable-flavor number scheme).
In the FFNS, which is valid at kinematics comparable to the heavy-quark mass and below (\(\pT \lesssim \mQ\)), the heavy quark is treated as a massive particle, whereas the lighter quarks are considered massless partons.
This takes into account \((\mQ/\pT)^n\) power corrections.
Collinear logarithms are finite because of the non-zero heavy-quark mass.
On the contrary, in the ZM-VFNS, the heavy quark is treated as another parton, which, in the same way as the light quarks, activates at its respective mass threshold.
This means the ZM-VFNS is valid in the kinematical region where the heavy-quark mass can be neglected (\(\pT \gg \mQ\)).
Power corrections are not present because the heavy-quark mass is zero.
The latter also causes the collinear logarithms to diverge, so they are renormalized in the usual \(\overline{\text{MS}}\) scheme.

If one now tries to apply the FFNS in the region where the heavy-quark mass is negligible (\(\pT \gg \mQ\)), the collinear logarithms get large, which spoils perturbative convergence.
On the other hand, in the kinematic region comparable to the heavy-quark mass (\(\pT \lesssim \mQ\)), considering the heavy quark as massless neglects the power-correction terms.
The solution is the GM-VFNS \cite{kniehl_inclusive_2005,kniehl_collinear_2005}, which takes the power-correction terms into account and involves additional subtraction terms that cancel the collinear logarithms.
Thus it is valid over the whole kinematic range, and in principle reduces to the FFNS and ZM-VFNS in their respective regions of applicability.

\section{Introduction to Fast-Convolution Grids with PineAPPL}

Gridding libraries like FastNLO \cite{kluge_fastnlo_2007}, APPLgrid \cite{carli_posteriori_2010} and PineAPPL \cite{carrazza_pineappl_2020} solve the issue of sufficiently performant predictions for PDF fits.
They store the hard-scattering cross-section of a process in PDF- and \(\alphas\)-independent interpolation grids.
In this way, the PDF- and \(\alphas\)-functions can be swapped a-posteriori, i.e.~after running the MC generator, which is useful when comparing different PDF sets or when calculating PDF uncertainties.
Since PDF sets may contain \(\mathcal{O}(100)\) PDF members (e.g.~200 for nNNPDF3.0 \cite{abdul_khalek_nnnpdf30_2022}), gridding libraries can speed up calculating the PDF uncertainties of a prediction by a factor of \(\mathcal{O}(100)\) for such PDF sets.

In the following, we sketch the process of constructing a grid in the momentum fraction \(x\), e.g.~in the case of a static scale \(Q\).
The idea behind producing such an interpolation grid is to expand a PDF \(\mathop{f}(x)\) in the Lagrange basis polynomials \(\mathop{L_i}(x)\),
\begin{equation} \label{eq:lagrange-pdf}
    \mathop{f}(x) = \sum_i \, f_i \mathop{L_i}(x) \, ,
\end{equation}
which can then be plugged into the factorization theorem for the process (here considering one PDF and leaving out \(\alphas\) for simplicity):
\begin{equation} \label{eq:lagrange-factorization}
    \d \sigma = \int \d x \mathop{f}(x) \mathop{\d \hat{\sigma}}(x) \overset{\eqref{eq:lagrange-pdf}}{=} \sum_i \, f_i \int \d x \mathop{L_i}(x) \mathop{\d \hat{\sigma}}(x) \eqcolon \sum_i \, f_i \, \d \sigma_i \, .
\end{equation}
When doing MC integration, the \(\d \sigma_i\) are the MC weights, calculated by randomly sampling points \(x^{(j)}\) from the phase space:
\begin{equation} \label{eq:lagrange-weights}
    \d \sigma_i \coloneq \int \d x \mathop{L_i}(x) \mathop{\d \hat{\sigma}}(x) \overset{\text{MC}}{=} \frac{1}{N} \sum_{j=1}^N \mathop{L_i}(x^{(j)}) \mathop{\d \hat{\sigma}}(x^{(j)}) \, .
\end{equation}
While running the MC, the \(\d \hat{\sigma}/N\) have to be passed to the gridding library to build the grid.
This takes as much time as running the MC integrator itself, with a small overhead from producing the grid, which is usually saved to disk after the integrator finishes.

Afterward, the prediction can be calculated by the gridding library from the grid file, which is again read from the disk and convolved with a PDF according to \cref{eq:lagrange-factorization}. This can be regarded as instantaneous for grid files of reasonable size.

\section{Results}
We use the GM-VFNS Fortran code written for \cite{kniehl_inclusive_2005,kniehl_collinear_2005}, last used in \cite{benzke_b-meson_2019}, extended by a custom PineAPPL interface.
In the following, we will present some results.
All of the results are given as double-differential cross sections in transverse momentum and rapidity of the D- or B-meson in the final state, with the bin limits taken from an experimental dataset given in the title of the figure.

\begin{wrapfigure}[16]{r}{0.48\textwidth}
    \vspace{-8pt}
    \centerfloat{\includegraphics[width=0.5\textwidth]{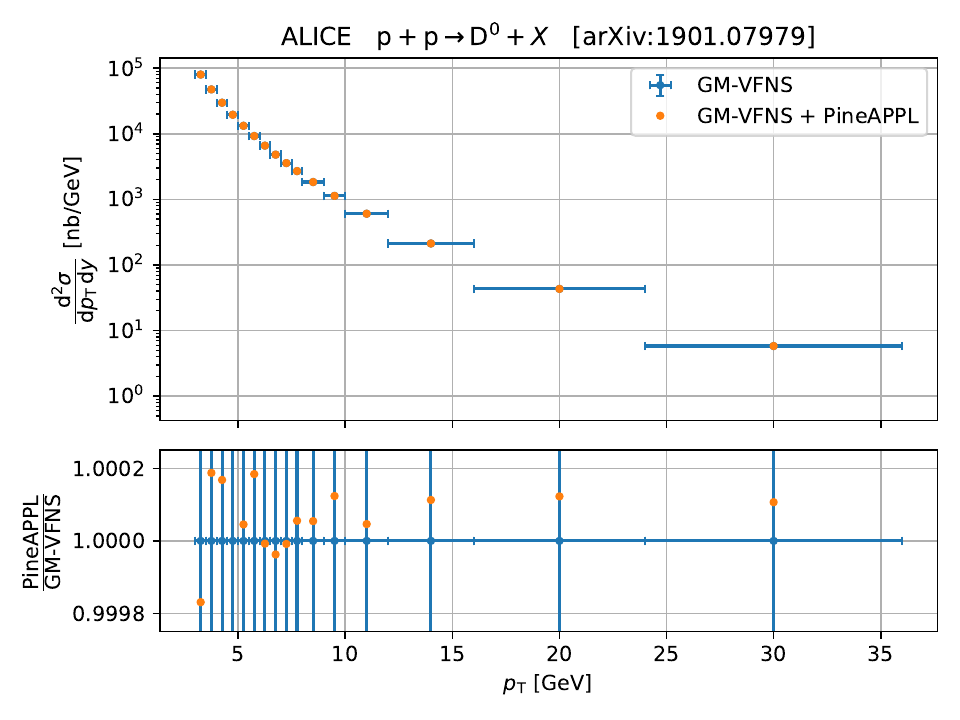}}
    \vspace{-8pt}
    \caption{GM-VFNS prediction obtained directly from the MC (\enquote{GMVFNS}) vs. prediction obtained from the grid (\enquote{GMVFNS+PineAPPL}). Displayed are the statistical MC uncertainties. The grid agrees with the MC prediction within less than one permille.}
    \label{fig:grid-vs-prediction}
\end{wrapfigure}
Unless stated otherwise, we use the proton and lead PDFs of nCTEQ15 \cite{kovarik_ncteq15_2016} (\texttt{nCTEQ15\_1\_1} and \texttt{nCTEQ15FullNuc\_208\_82}, respectively).
All scales are chosen as \(\sqrt{\pT^2 + (2m)^2}\), where the factor of 2 is justified by the production threshold of the final state, since in the FFNS cross section always a \(Q\widebar{Q}\) pair must be produced.
We do not consider fine-tuning the coefficient of the scale here, which was shown to improve the low-\(\pT\) behavior \cite{kniehl_inclusive_2015,benzke_b-meson_2019}, instead we only consider the data above a cut of \(\pT = \qty{3}{GeV}\).
In all results we set the charm and bottom mass to \qty{1.3}{\GeV} and \qty{4.5}{\GeV}. Furthermore, as fragmentation functions we use KKKS08 \cite{kneesch_charmed-meson_2008} and BKK06 \cite{kniehl_finite-mass_2008}.

\begin{wrapfigure}[17]{l}{0.48\textwidth}
    \vspace{-8pt}
    \centerfloat{\includegraphics[width=0.5\textwidth]{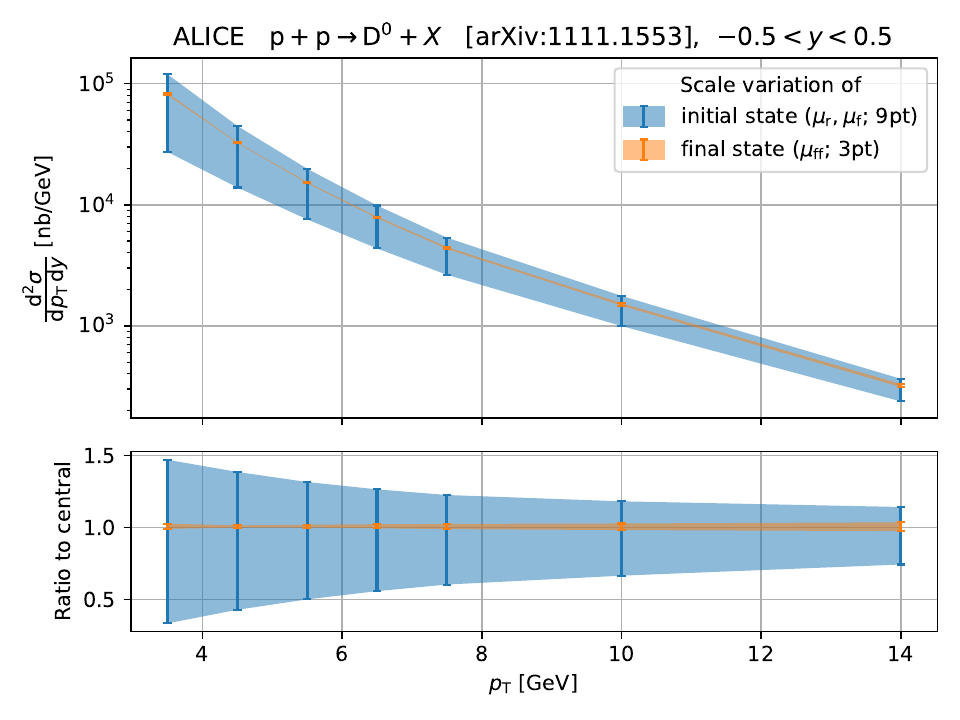}}
    \vspace{-8pt}
    \caption{Renormalization and initial-state factorization scale dependence (\enquote{initial state}) vs. fragmentation scale dependence (\enquote{final state}). The fragmentation scale dependence is negligible relative to the renormalization and initial-state factorization scale.}
    \label{fig:scale-initial-final}
\end{wrapfigure}
First, we check the agreement of the PineAPPL grids with the GM-VFNS predictions in \cref{fig:grid-vs-prediction}.
The input for the transverse-momentum and rapidity bins as well as the cms energy \(\sqrt{s} = \qty{5.02}{\TeV}\) are taken from the ALICE dataset \cite{alice_collaboration_measurement_2019-1} displayed in the title of the figure.
Compared are the predictions from the MC integrator (blue) and from the PineAPPL grid (orange).
Both were produced in the same run (using a VEGAS integrator, utilizing 30000 calls in 4 iterations), after which the grid was convolved with the \texttt{nCTEQ15\_1\_1} PDF for comparison.
The grid prediction lies within the statistical errors (vertical bars) of the pure MC prediction and agrees with its central value within one permille.
Estimating the PDF and scale uncertainties as \(\qty{10}{\%}\), which usually get much larger going to lower \(\pT\), this deviation of the grid is negligible.
Since the grid is produced with the same weights as the pure MC predictions, and thus the same sampling is used for both, the grid precision is not dependent on the kinematics of the run.
This was also verified for the kinematic input of other datasets.
However, disagreement can be found in numerically unstable regions of phase space, e.g.~the low-\(\pT\) region, which is not considered here.
Thus we conclude that the grids reproduce the GM-VFNS predictions sufficiently for any practical applications.

Next, we examine the dependence of the predictions on the final-state factorization scale, which we also call the \enquote{fragmentation} scale in the following.
Since the fragmentation function is \enquote{baked into} the grid and cannot be varied a-posteriori, the impact of this limitation should be assessed.
To do this, in \cref{fig:scale-initial-final} we vary separately the fragmentation scale (orange) and the renormalization and initial-state factorization scales (blue) by the factors \(\{0.5, 1, 2\}\) each.
The bins are again chosen with respect to an {ALICE} dataset \cite{alice_collaboration_measurement_2012} of cms energy \(\sqrt{s} = \qty{7}{\TeV}\), but other kinematic configurations not shown here were checked as well.
In the lower- and mid-\(\pT\) regions, the fragmentation scale uncertainty computes to less than \qty{10}{\%} of the renormalization and initial-state factorization scale uncertainties.
These are still much more dominant in the last bin, even though the proportion of the fragmentation scale uncertainty increases with increasing \(\pT\).
Therefore, we estimate the uncertainty of baking the fragmentation into the grid as negligible.

\begin{wrapfigure}[13]{r}{0.48\textwidth}
    \vspace{-10pt}
    \centerfloat{\includegraphics[width=0.5\textwidth]{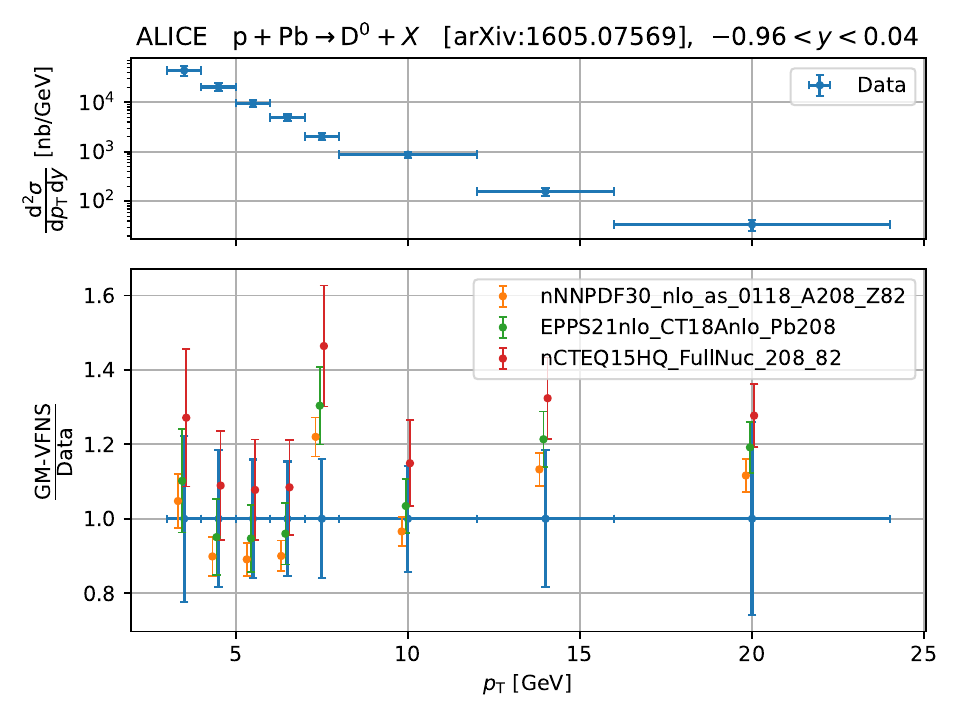}}
    \vspace{-10pt}
    \caption{GM-VFNS prediction with PDF uncertainties.}
    \label{fig:pdf-unc}
\end{wrapfigure}
In \cref{fig:pdf-unc} we show the PDF uncertainties of the GM-VFNS predictions for an ALICE dataset \cite{alice_collaboration_d-meson_2016} with the nCTEQ15HQ \cite{duwentaster_impact_2022}, EPPS21 \cite{eskola_epps21_2022} and nNNPDF3.0 \cite{abdul_khalek_nnnpdf30_2022} PDF sets.
The predictions seem to increase consistently in the order nNNPDF, EPPS21, nCTEQ15HQ, which also resembles the order in which the uncertainties increase.
Except in the bin \(\pT \in [7, 8]\,\unit{\GeV}\), the PDF uncertainty intervals overlap with the data intervals.
The PDF sets shown in \cref{fig:pdf-unc} include 347 PDF members in total, such that gridding reduces the computation time by about a factor of 347.

\begin{figure}[ht!]
    \centering
    \begin{subfigure}[b]{0.49\textwidth}
        \centering
        \includegraphics[width=\textwidth]{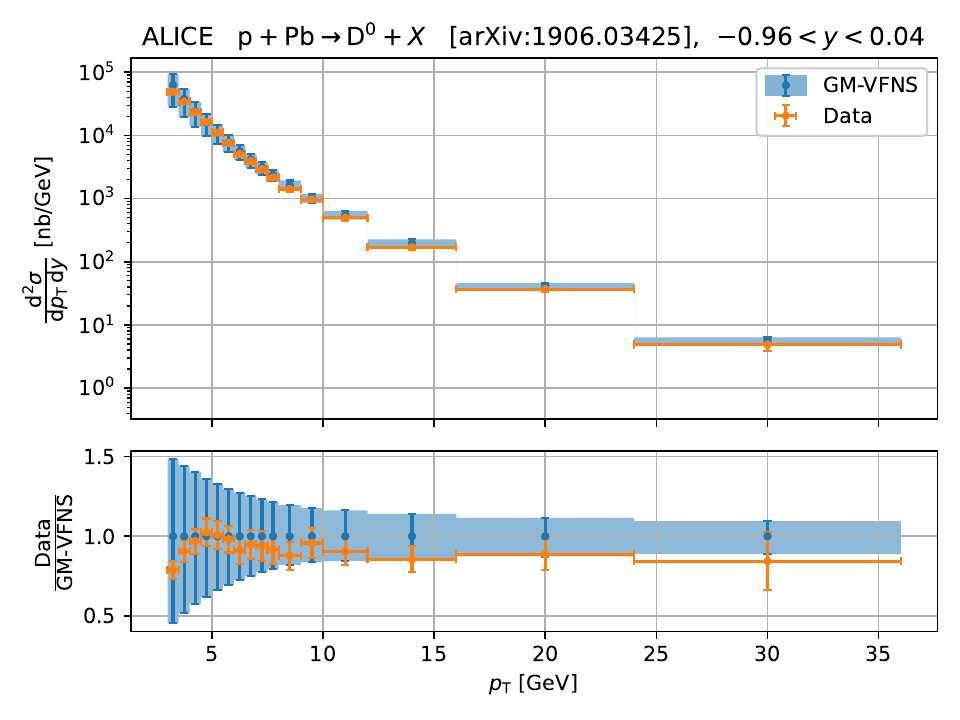}
        \caption{}
        \label{fig:prediction-vs-data-alice}
    \end{subfigure}
    \hfill
    \begin{subfigure}[b]{0.49\textwidth}
        \centering
        \includegraphics[width=\textwidth]{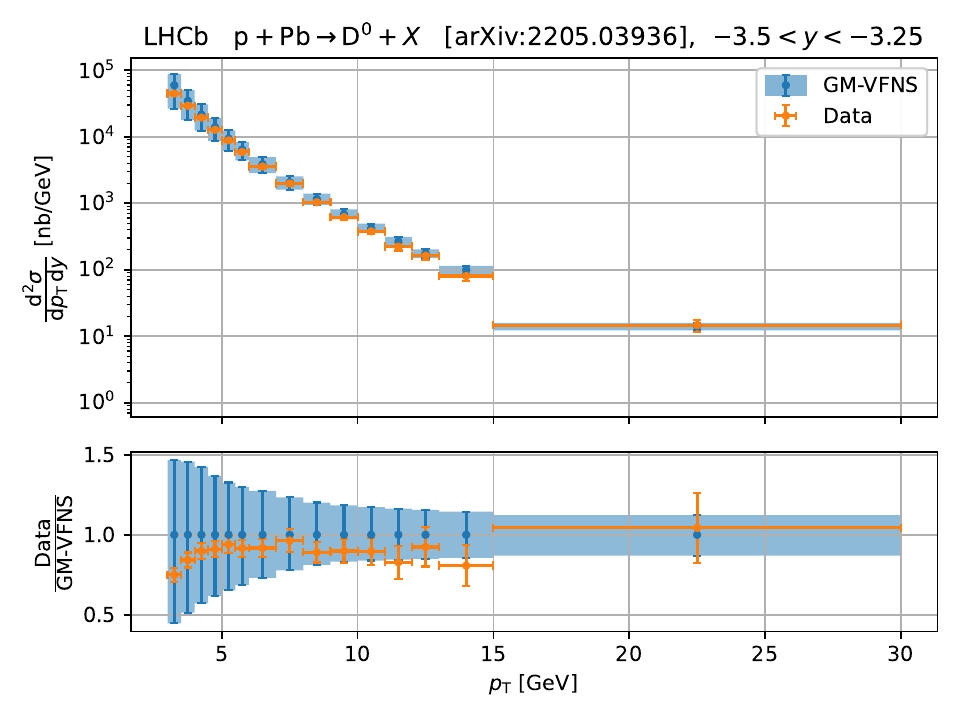}
        \caption{}
        \label{fig:prediction-vs-data-lhcb}
    \end{subfigure}
    \hfill
    \begin{subfigure}[b]{0.49\textwidth}
        \centering
        \includegraphics[width=\textwidth]{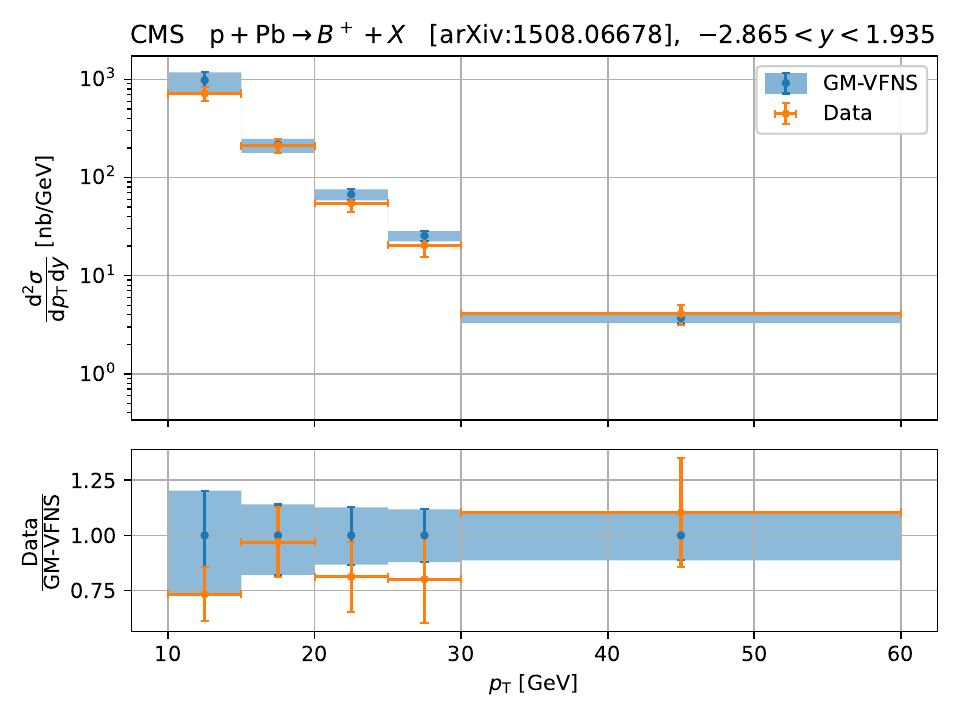}
        \caption{}
        \label{fig:prediction-vs-data-cms}
    \end{subfigure}
    \caption{GM-VFNS predictions~vs.~data. The displayed theory uncertainties are obtained from 7-point scale variations of the renormalization and the initial-state factorization scales.}
    \label{fig:prediction-vs-data}
\end{figure}

Lastly, in \cref{fig:prediction-vs-data} we compare GM-VFNS predictions to data for \(\mathrm{D}^0\)- and \(\mathrm{B}^{\pm}\)-production in p-Pb collisions from ALICE \cite{alice_collaboration_measurement_2019}, LHCb \cite{lhcb_collaboration_measurement_2023} and CMS \cite{cms_collaboration_study_2016}.
The theory uncertainties are 7-point scale variations of the renormalization and initial-state factorization scale by the factors \(\{0.5, 1, 2\}\) such that there is never a factor of 2 between the scales.
The GM-VFNS predictions agree with the data within the scale uncertainties.

\section{Conclusion}
To make GM-VFNS predictions for heavy-quark hadroproduction usable for PDF fits, we implemented a PineAPPL interface for an existing GM-VFNS code.
We compared the grids to the pure-MC predictions, which agree within one permille.
Additionally, we showed agreement between GM-VFNS predictions and data within scale and PDF uncertainties.
Computation of the PDF uncertainties was sped up by 2 orders of magnitude due to calculating them with a grid.
We plan to publish GM-VFNS heavy-quark hadroproduction grids as part of a future publication.

\section{Acknowledgements}
We thank our nCTEQ colleagues for useful discussions and valuable feedback. Work in Münster was funded by the Deutsche Forschungsgemeinschaft (DFG) through the Research Training Group GRK 2149.

\bibliography{references}

\end{document}